\documentclass{ws-procs10x7}
\begin{document}

\title{Summary of ICHEP 2004}

\author{John Ellis}

\address{Theory Division, CERN, CH-1211 Geneva 23, Switzerland}

\twocolumn[\maketitle\abstract{Aspects of ICHEP 2004 are summarized from a
theoretical point of view. QCD works, new NNLO calculations are becoming
available and new string calculational tools are emerging, but no
conclusions can yet be drawn about the discovery of the quark-gluon plasma
or pentaquarks.  The small upward shift in the experimental value of $m_t$
raises somewhat the central value of the Higgs mass extracted from a
global electroweak fit, and the CKM model describes well the data from the
B factories. The Super-Kamiokande, KamLAND and K2K experiments have
evidence for oscillation dips in their neutrino data. Little Higgs models
are interesting alternatives to low-energy supersymmetry for stabilizing
the electroweak scale. Convincing experimental evidence for dark matter
particles is still lacking. The LHC is on its way, the technology choice
clarifies the roadmap for the ILC, and a multi-TeV CLIC would also have
rich physics agenda.

\begin{center}
hep-ph/0409360 $\; \; \;$ ~~~~~~~~~~CERN-PH-TH/2004-167
\end{center}
}]

\section{Introduction}

Summarizing this conference with its wealth of new results is an
interesting challenge, and I apologize in advance to those whose work I
neglect or under-emphasize.

A good place to start my task is the basic Lagrangian underlying
particle
physics~\cite{Barbieri}:
\begin{eqnarray}
{\cal L} & = & - \frac{1}{4} \, F^a_{\mu \nu} + i \bar{\psi} \not{D} \psi
+ \psi_i \lambda_{ij} \psi_j h + h.c. \nonumber \\
& + &|D_{\mu} h|^2 - V(h) \nonumber \\
& + & (\frac{1}{M}\, L_i \lambda^{\nu}_{ij} L_j h^2 \, {\rm and/or} \,
L_i \lambda^{\nu}_{ij} N_j h + h.c.) \nonumber
\end{eqnarray}

The first two terms constitute the gauge sector of the theory, the next two
terms describe quark flavour physics, the
next two are alternative and complementary contributions to neutrino
masses, and the last two terms represent the
electroweak symmetry-breaking sector of the Standard Model. The following
sections of this talk discuss these various
sectors in turn. Subsequently, sections are devoted respectively to string
theory, the connection between particle
physics and cosmology, and the prospects for future accelerators. Finally,
this conference summary is itself
summarized by a few concluding remarks.

\section{The Gauge Sector}

\subsection{QCD}

The challenge in QCD is no longer to test the theory, but rather to
understand and calculate it better~\cite{Stirling}.
Perturbative calculations are most reliable in the high-energy regime,
where they provide the essential baselines
for searches for new physics beyond the Standard Model. The high
temperatures and pressures attained in
relativistic heavy-ion collisions may eventually offer us another
relatively simple playground, and the lattice is
an increasingly accurate tool for non-perturbative calculations, but hadron
spectroscopy does not cease to pose
important challenges to our qualitative understanding of QCD.

As we heard here, the latest data from both the FNAL 
Tevatron~\cite{Lucchesi} and HERA~\cite{Klein} at
DESY are in good agreement with
perturbative QCD calculations. This is true, in particular, for the total
jet cross sections (after adjusting the parton distribution functions) and 
for heavy-flavour
production. In the latter case, previous discrepancies between theory and
experiment have dissipated with the
advent of new calculations~\cite{newcalx} and 
measurements~\cite{Lucchesi,Klein}. This is 
reassuring for the
LHC, which will depend on Tevatron and HERA
inputs in its searches for new physics. Several new measurements of
$\alpha_s$ were reported here, in particular
from HERA~\cite{Klein}. The current world average value of $\alpha_s (m_Z)$ 
is $0.1182
\pm 0.0027$~\cite{Bethke}, which (reassuringly) has
changed little from the corresponding average a couple of years ago.

Significant progress has been reported here in the drive towards greater
precision in QCD calculations~\cite{Stirling}. The full
NNLO expressions for the quark and gluon splitting functions are now 
known~\cite{Moch}, and cover several pages! They lie
comfortably within the range expected from previous incomplete
calculations. They open the way to a new era of
high-precision QCD, which has already been inaugurated by comparisons of W
and Z production data from the Tevatron
collider with NNLO cross sections~\cite{Stirling} - these might provide a 
useful tool for
measuring the LHC luminosity. Important
parts of the NNLO calculations of hadronic jet cross sections have also
been completed, and are likely to be
complete by the time the LHC is switched on, though there are still
important technical issues in the infra-red cancellations needed for
the NNLO calculations for multi-jet production.

\begin{figure}[htb]
\begin{center}
\includegraphics[width=.45\textwidth]{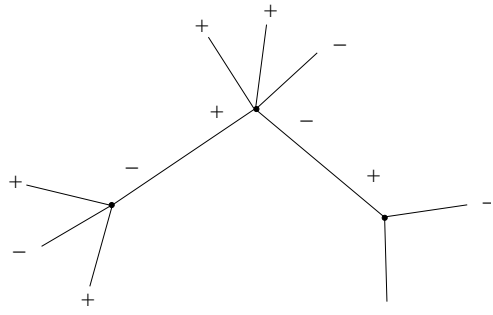}
\caption{Illustration how MHV multigluon amplitudes can be combined 
to yield tree-level non-MHV amplitudes~\protect\cite{CSW}. The $\pm$ signs 
indicate the gluon helicities.}
\label{fig:MHV} 
\end{center}
\end{figure}

A very exciting theoretical development has been the proposal of a new and
powerful string approach to QCD calculations~\cite{CSW}. It uses the
simple amplitudes for maximal helicity-violating (MHV) multi-gluon
processes~\cite{Kiwi} as effective vertices in a new graphical approach
based on scalar field theory. The MHV amplitudes are combined with scalar
propagators to calculate tree-level non-MHV amplitudes for both quarks and
gluons, as illustrated in Fig.~\ref{fig:MHV}, and may also be used to
calculate loop diagrams~\cite{CSW}. This method is dramatically simpler
than conventional techniques, providing compact outputs expressed in terms
of familiar spinor products. There are high hopes of using this technique
for phenomenology~\cite{Khoze}, with calculations of multijet cross
sections underway for the LHC.

Another `hot' topic at the moment is the QCD phase diagram, which is
expected to exhibit a colour-superconductor
phase at high baryon density and a quark-gluon-plasma (QGP) phase at high
temperatures, that may be explored in
relativistic heavy-ion collisions~\cite{Dunlop}. These phases are not 
expected to be separated by a strong first-order phase 
transition, though there may be an interesting critical point to explore, 
if one can devise observable signatures of it. Thermodynamics 
should allow reliable calculations to be made for the QGP phase,
but our ability to confront them with experiment is hampered by the facts
that thermalization may not be complete, that the QGP is expected to be 
strongly-interacting~\cite{QGP},
and that many probes involve hadrons produced after the QGP has cooled.
Nevertheless, the measured abundances of the
many particle species  are well described by a simple thermodynamic model,
with a freeze-out temperature $T$
similar to that predicted by lattice calculations, and a baryonic chemical
potential that decreases with increasing
collision energy. However, is it possible that these abundances may just 
be reflecting phase space and statistics?

Considerable interest has arisen from RHIC measurements of the elliptic
flow variable $v_2$~\cite{Dunlop}, which reflects the
shape of the interaction region. In peripheral collisions, this is expected
to start with an almond shape. It
should then expand more rapidly in the direction of the minor axis, where
the pressure is greater, so that the
shape becomes more spherical at later times as the interaction region
cools. This expected behaviour is seen in the
transverse momentum spectra, which exhibit higher $v_2$ for the
higher-$p_T$ particles that are presumably more
likely to have been produced earlier. Moreover, the elliptic flows for
different hadrons scale approximately with
the numbers of quarks they contain,
\begin{equation}
v_2^M(p_T) \sim 2 v_2^q (p_T/2),
v_2^B(p_T) \sim 3v_2^q(p_T/3),
\end{equation}
in agreement with the idea that the values of $v_2$ originate with their
constituents during a QGP phase.
The measurements are reproduced well by a hydrodynamic model using a
simplified quark-gluon equation of state.

However, simple hydrodynamic models fail to reproduce the source size
inferred from HBT interference measurements~\cite{Dunlop}. If some 
quark-gluon fluid is
being formed, it seems to resemble a liquid rather than an ideal gas, and
complications such as strong interactions and viscosity may need to be
incorporated in modelling it.

\begin{figure}[htb]
\begin{center}
\includegraphics[width=.45\textwidth]{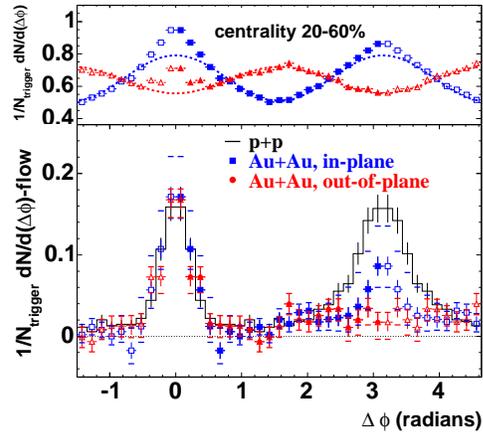}
\caption{
The azimuthal distributions of high-$p_T$ particles in $Au+Au$ collisions 
at RHIC (bottom panel) 
differ from the standard pattern of elliptic flow (top 
panel)~\protect\cite{STAR}. The 
suppression compared with $p+p$ collisions (shown) and $D+Au$ collisions 
(not shown) is interpreted as the quenching of jets by final-state 
interactions in a hot and dense medium. The opposite-side suppression is 
greater out of the reaction plane than in it, as expected because the 
pressure and density should be greater in that 
direction~\protect\cite{Dunlop}.} \label{fig:quench} \end{center} 
\end{figure}

The poster-child for QGP production at RHIC may be the observation of jet
quenching~\cite{Dunlop}, illustrated in Fig.~\ref{fig:quench}~\cite{STAR}. 
This occurs 
in central $Au+Au$
(but not $D+Au$ or minimum-bias $p+p$) collisions on the side opposite a
transverse jet, and is stronger out of the
overall event plane, where the initial density is thought to be 
higher. The QGP
interpretation is that the opposite-side parton
loses its transverse energy in collisions with coloured partons in the QGP.
This jet quenching is accompanied by an
increased $\gamma / \pi^0$ ratio at larger $p_T$, and softening of the jet
fragmentation function.

Are these observations, as well as others including the enhancement of
low-mass pairs and the suppression of
$J/\psi$ particles observed in earlier relativistic heavy-ion collisions at
the CERN SPS, sufficient to claim
discovery of the expected QGP? RHIC data are still described by a patchwork
of theoretical approaches, with model
parameters often adjusted independently for different observables.
Moreover, some observables, such as particle
ratios, may have alternative explanations, and others, such as the HBT
radii are not yet reproduced well by simple QGP
calculations. The parton energy loss apparently observed in jet quenching
is a very promising development, and it
would be interesting to hear from RHIC about the $J/\psi$ suppression
already observed at the CERN SPS. In the
absence of a `smoking gun' for a compelling QGP claim, I share the view of
the rapporteur here~\cite{Dunlop} that we still need more
quantitative estimates of the theoretical uncertainties.

The final QCD topic to discuss is hadron spectroscopy,
which is experiencing an experimental
renaissance at the moment~\cite{Jin}. There has been much discussion here 
whether
pentaquarks or other exotic hadrons exist.
The quark descriptions of the $D_{sJ}(2317)$ and $D_{sJ}(2460)$ have also
been debated~\cite{Bardeen}, as has the existence of the
$D_{sJ}(2632)$ state reported by SELEX~\cite{SELEX}. The quark description 
of the
$X(3872)$ discovered by BELLE is also intriguing, as is the fate of
the 12$\%$ rule in $\psi '$ decay investigated by BES and the
interpretations of other threshold states they have
reported here. In the time available, I shall concentrate on pentaquarks,
the $X(3872)$ and some remarks about
threshold states.

\begin{figure}[htb]
\begin{center}
\includegraphics[width=.45\textwidth]{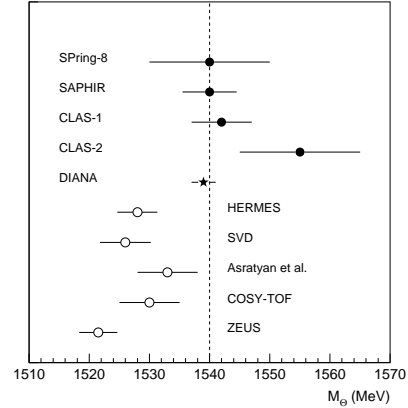}
\caption{
The mass of the $\Theta^+(1540)$ measured in $nK^+$  final states (solid 
circles) tends to be larger than in $pK^0_s$  final
states (open circles)~\protect\cite{ZC,Close,Jin}. The starred 
measurement was made in an experiment connecting $nK^+ \to pK^0_s$.}
\label{fig:ZC}
\end{center}
\end{figure}

Sightings of the $\Theta^+(1540)$ pentaquark candidate were reported here
by HERMES~\cite{HERMES} and ZEUS~\cite{ZEUS}, as well as the
observation of a $\Theta_c(3095)$ by H1~\cite{H1}. In addition to these, 
there have
also been many other observations of the
$\Theta^+(1540)$~\cite{Jin}, and one report of an exotic $\Xi^{--}(1862)$ 
baryon~\cite{NA49}. One
puzzle in the $\Theta^+(1540)$
observations has been that the masses and decay widths have varied between
experiments~\cite{Close}. Generally 
speaking,
observations of the
$\Theta^+(1540)$ in the $nK^+$  final state tend to give higher masses 
than
observations in the $pK^0_s$  final
state, as seen in Fig.~\ref{fig:ZC}. Also, the two positive experiments 
reporting observations here yield
non-zero decay widths: $\Gamma_{\theta}
= 17 \pm 9 \pm 2$ MeV~\cite{HERMES} and
$\Gamma_{\theta} =
8 \pm 4$ MeV~\cite{ZEUS}, whereas partial-wave analyses of $K^+N$ 
scattering
indicate an upper limit $\Gamma_{\theta} <
1$ MeV~\cite{PW}.

Several other experiments (including a number reported here) have not seen
these candidate pentaquark
states~\cite{Jin}. How can one assess the significance of these negative 
experiments?
One strategy is to compare the obtained
upper limit with the trend line of the measured rates for the production of
other, well-established baryons. The
(upper bound on the) yield of the $\Theta^+(1540)$ is often compared with
that of the $\Lambda(1520)$. However, as
discussed here by the BABAR collaboration~\cite{BABAR}, the $\Lambda(1520)$ 
yield is
anomalously high, compared with trend line.
Nevertheless, the upper bound on $\Theta^+(1540)$ production is
significantly lower than this trend line, as is the
BABAR upper limit on $\Xi^{--}(1862)$ production. Generally speaking, most
positive results for $\Theta^+(1540)$
are at relatively low energies, and not at high energies~\cite{Danilov}, 
suggesting that
exotic baryons may have production
mechanisms that are different from those of conventional baryons. For 
example, their 
photo- and electroproduction may be facilitated by the ${\bar s} s$ 
component of the photon.

The interpretation of the $\Theta^+(1540)$ could be very 
interesting~\cite{Close,Lipkin} --- {\bf if it exists!} In order to 
accommodate
it, na\"{i}ve non-relativistic quark models would need epicycles such as
di- or triquarks~\cite{JW,KL}, and the ground state
would involve some $P$-wave configuration, an unfamiliar option within the
quark model. However, something very
much like the $\Theta^+(1540)$, in both mass and narrow width, was
predicted~\cite{DPP} in the chiral soliton model~\cite{Skyrme}, which is
based on the idea that quarks are intrinsically very light --- a few MeV
for the up and down and $\sim 100$ MeV for
the strange --- and that baryons are topological clouds effectively
composed of very many quarks. In addition to
fitting the data on the $\Theta^+(1540)$ and the $\Xi^{--}(1862)$, the
chiral soliton model also predicts other
exotic states, filling out an antidecuplet of $SU(3)$, as well as other
nearby $27-$ and $35$-dimensional
representations~\cite{more}.

The existence of the
$\Theta^+(1540)$ still lacks conclusive confirmation --- even a single
high-statistics, high-significance experiment
would suffice. If it does exist, measurements of the $\Theta^+$ spin and
parity would distinguish between the chiral
soliton and rival quark models. The stakes are high: {\bf if they exist},
the $\Theta^+$, $\Xi^{--}$ and $\Theta_c$
may take us beyond the na\"{i}ve quark model.

The next spectroscopic surprise that I discuss is the $X(3872)$
state~\cite{Jin} discovered by Belle~\cite{Belle} and confirmed by
CDF~\cite{CDF} and BABAR~\cite{BABAR2}. Here the theoretical debate is
whether it is a (displaced)  charmonium state, which would need unnatural
spin-parity such as $1^{++}$ in order to avert rapid decay into
$D\bar{D}$, or, in view of its near-degeneracy with the $D^0 D^{*0}$
threshold, whether it should better be regarded as a $D^0 D^{*0}$
`molecular state'~\cite{Close}. The suspicion is that the $\pi \pi$ pair
in the $\pi \pi J/\psi$ mode first observed may be in a $\rho$ state, in
which case the recent Belle report of $X(3872) \to \omega J/\psi$
decay~\cite{Belle2} would be very interesting.  It would be evidence for
isospin violation in $X(3872)$ decay, which would be a signature of a $D^0
D^{*0}$ admixture in its wave function.

There have been several other reports here of hadronic threshold states,
from BES~\cite{BES} and Belle~\cite{Belle3} in particular. The most
impressive is the $\bar{p}p$ state with a mass of $M =
1859^{+~3+~5}_{-10-25}$ MeV and a decay width $\Gamma < 30$ 
MeV~\cite{BES}, 
but
other states have also been reported in the $\bar{p}\Lambda$,
$\bar{p}\Lambda_c$, $\bar{K}\Lambda$, $\pi \pi$~\cite{BESpipi} and
$\pi K$ combinations~\cite{Jin}. These might be candidates for tetra- or 
even
sextaquark states, but an alternative would be
that the quark description is not optimal for discussing such
non-relativistic hadronic combinations~\cite{Close}. Perhaps they
should rather be described in terms of hadronic physics in a first
approximation, as are nuclei.

\subsection{Electroweak Physics}

The electroweak sector of the Standard Model continues to withstand
experimental attacks~\cite{Teubert}, despite some points of
hot discussion such as the comparison between low- and high-energy data.
Measurements of parity violation in atomic
physics and M{\o}ller scattering are compatible with the expected running
of $\alpha_{em}(Q^2)$, but the NuTeV
measurement of $sin^2_{W}$ is offset by about $3 \sigma$~\cite{NuTeV}. 
Improved
electroweak radiative corrections as well as an
asymmetry in the strange sea and a difference: $\bar{u} \not= \bar{d}$
might each account for about $1 \sigma$ of
this discrepancy~\cite{Teubert}. Light will be cast on the NuTeV 
measurement of
$sin^2_{W}$ by the forthcoming NOMAD result~\cite{NOMAD}, so I
follow the rapporteur~\cite{Teubert} in leaving NuTeV out of the global 
electroweak fit
for the time being.

The big news in the electroweak sector has been the new value of the
top-quark mass reported by D{\O}~\cite{D0}, on the basis of a re-analysis
of their Run 1 data: $m_{t} = 179.0 \pm 5.1$ GeV which, when combined with
the previous CDF measurement, yields a world average $m_{t} = 178.0 \pm
4.3$ GeV. There are good prospects for further reducing the experimental
error using Run 2 data from both D{\O} and CDF~\cite{Denisov}.

\begin{figure}[htb]
\begin{center}
\includegraphics[width=.45\textwidth]{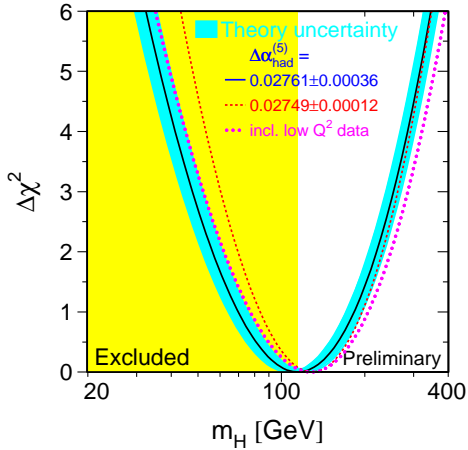}
\caption{
The $\chi^2$ function in the latest global fit to 
$m_H$~\protect\cite{Teubert,LEPEWWG}, incorporating the 
new D{\O} value of $m_t$~\protect\cite{Denisov,D0}, the latest 
theoretical 
calculations~\protect\cite{Weiglein} and their uncertainties (blue band). 
Also shown are the effects of incorporating low-energy data or a 
theory-driven value for 
$\alpha_{em}(m_Z)$, and the lower limit from the direct search at LEP 
(yellow shading)~\protect\cite{LEPHWG}.} 
\label{fig:blueband}
\end{center}
\end{figure}

The D{\O} result is important news, in particular because of its
implications for the Higgs mass predicted in the
global electroweak fit~\cite{Teubert,LEPEWWG}. The new value of $m_t$ 
increases the prediction for
$m_H$ by $\sim 20$ GeV and new 2-loop terms
in
$m_W$ and $\sin^2_W$~\cite{Weiglein}, as well as other theoretical 
improvements, increase
$m_H$ by $\sim 6$ GeV, yielding
$$
m_H = 114^{+69}_{-45} \quad {\rm GeV}
$$
as the present best estimate using an experiment-driven~\cite{BESetal} 
value of
$\alpha_{em}(m_Z)$: see Fig.~\ref{fig:blueband}. The central
value of $m_H$ would be increased by $\sim 15$ GeV if a theory-driven value
of $\alpha_{em}(m_Z)$ with a smaller error
were chosen~\cite{Teubert}. The largest pull on the global electroweak fit 
comes from the
forward-backward asymmetry for $b$ quarks, and
heavy-flavour measurements of $\sin^2_W$ tend to yield values different
from leptonic measurements and $m_W$,
favouring larger values of $m_H$. However, the overall quality of the
global electroweak fit is good: $\chi^2 = 15.8$
for 13 degrees of freedom, corresponding to an overall probability of 26\%,
and yields a 95\% upper limit $m_H  <
260$ GeV~\cite{Teubert}.

The central fit value of $m_H$ now {\it coincides} with the lower limit 
from
direct searches at LEP~\cite{LEPHWG}, and the hunt for the
Higgs at hadron colliders is very much on. The latest CDF upper limits on
Higgs production are almost within an
order of magnitude of the cross section expected in the Standard Model for
$m_H$ between 110 and 180 GeV~\cite{Denisov}, and the
Tevatron collider has a window of opportunity before the LHC experiments
start taking data. Once they start, even a
small amount of running might suffice to discover the Higgs boson, once the
detectors are understood~\cite{Barr}.

\section{Quark Flavour Physics}

Before this conference, some of the big questions in quark flavour physics
were: Are the data on quark mixing well
described by the Cabibbo-Kobayashi-Maskawa (CKM) model? Are there
signatures of physics beyond the Standard Model?
If not, why is new physics flavour-blind? There are also some questions
about the relation to neutrino flavour
physics, namely why is neutrino mixing so different from quark mixing, and
is there some way they might be related?

The answer to the first quark flavour question seems still to be `yes'.
The niggling $2-\sigma$ discrepancy with CKM unitarity has evaporated with
a slew of new measurements of $V_{us}$ in $K$ decays that all give values
higher than the earlier PDG value~\cite{Patera}: there is no longer any
deficit of quark weak charge relative to muon decay. The previous CKM
unitarity `crisis' has disappeared~\cite{Patera}.

CKM has won again with the determination of $\sin 2 \beta$ in $b \to
\bar{c} c s$ decays such as $B \to J/\psi K_s$: the world average is now
$0.726 \pm 0.037$~\cite{betacc,Giorgi,Sakai}, which agrees fantastically
well with the prediction~\cite{Ali,Ligeti} based on measurements of $K$
decay and CP-conserving $B$ properties (and relying on lattice estimates
of weak matrix elements~\cite{Hashimoto}), confirming that CP violation is
`large'. However, this success with $\sin 2 \beta$ is only the beginning
of a paradigm change. CKM CP violation certainly present and large: the 
big issue is now to look for corrections to CKM,
rather than alternatives~\cite{Ligeti}. Further detailed tests of CKM in
clean processes are essential.

An important set of checks of the CKM model are provided by comparing the
value of $\sin 2 \beta$ extracted from $b
\to \bar{c} c s$ decays with the CP-violating asymmetries found in
processes dominated by $s$-penguin diagrams,
including $B \to \phi K^0, \eta' K_s, K^+K^-K_s, \pi^0 K_s, f^0K_s$ and
$\omega K_s$~\cite{betass,Sakai,Giorgi,Ali,Ligeti}. These measurements are
qualitatively consistent with $\sin2 \beta [cc]$, in the sense that they
exhibit large asymmetries of the same sign, but
there are some discrepancies in the details. Thanks to new data from 
Belle~\cite{Sakai},
in particular, $B \to \phi K_S$ has jumped
towards the Standard Model, but $B \to \eta' K_s$ is still $2.6 \sigma$
away from from $\sin2 \beta [cc]$. Overall, the
average of the
$s$-penguin processes now lies $3.5 \sigma$ away from $\sin2 \beta [cc]$.
However, none of the $s$-penguin processes is
as clean as $B \to J/ \psi K_s$, and one expects departures from $\sin2
\beta [cc]$ that are $O(15)\%$ for many modes~\cite{Ligeti}.
Moreover, there is no hint of any direct CP violation in any of these
modes. It would be premature to consider the
$s$-penguin decays to be a `smoking gun' for new physics.

\begin{figure}[htb]
\begin{center}
\includegraphics[width=.45\textwidth]{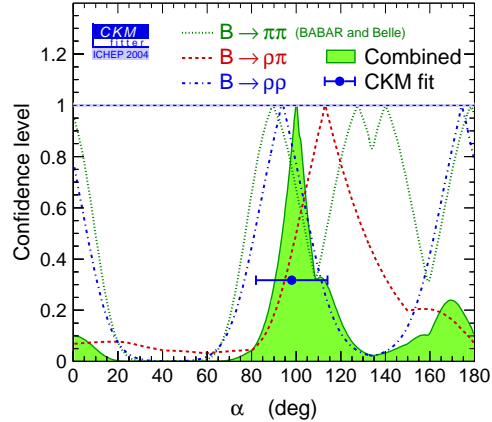}
\caption{
Prediction for the angle $\alpha$ from a global CKM 
fit~\protect\cite{Ligeti,CKMfitter}, compared with recent experimental 
determinations~\protect\cite{Giorgi,Sakai,pipi,rhopi,rhorho}.}
\label{fig:alpha}
\end{center}
\end{figure}

An important step forward in checking the CKM model has been achieved at
this conference with the measurement of $\alpha$ using $B \to \pi 
\pi$~\cite{pipi} ,
$\rho \pi$~\cite{rhopi} and $\rho \rho$~\cite{rhorho} 
decays~\cite{Giorgi,Sakai}, as shown in Fig.~\ref{fig:alpha}. The most 
important
contributors are $B \to \rho^+\rho^-$~\cite{Giorgi} (using the facts that
longitudinal $\rho$ polarization states dominate and that $B \to 
\rho^0\rho^0$ decay is suppressed, implying that penguin pollution
is small) and the Dalitz plot in $B \to \rho \pi$ decays. The world
average is now $\alpha = 100^{+12}_{-10}$ degrees, which is consistent
with, and more accurate than, the CKM prediction $\alpha = 98 \pm 16$
degrees. However, the direct determinations $\gamma = 
77^{+17}_{-19} \pm 13 \pm 11 {\rm (Belle)}, 88 \pm 41 \pm 
19 \pm 10 {\rm (BABAR)}$ degrees~\cite{gamma} are still some way from 
challenging the 
CKM model.

Another important result reported here by Belle and BABAR has been the
observation of direct CP violation in $B^0 \to K^+ \pi^-$ and conjugate
decays: $A_{CP}(K^+ \pi^-) = 0.114 \pm 0.020$, a 4.2-$\sigma$
effect~\cite{Kpi,Sakai,Giorgi}. This is only the second observation of
direct CP violation, following that in $K^0 \to \pi \pi$
decays~\cite{Patera}, and is the first in $B$ decays. It confirms that
these cannot be described by a superweak theory, and is consistent with
the CKM model. However, this measurement poses a problem for the
factorization scheme~\cite{factn}, and is somewhat different from the
asymmetry in $B^+$ decays: $A_{CP}(K^+ \pi^-) = 0.04 \pm 0.05 \pm 0.02$.
This difference might be laid at the door of some coalition of electroweak
penguins, non-perturbative effects and final-state
interactions~\cite{Ligeti}.

\begin{figure}[htb]
\begin{center}
\includegraphics[width=.45\textwidth]{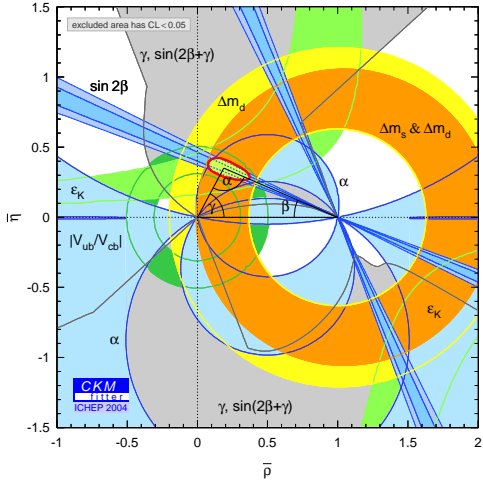}
\caption{
Measurements of CP violation in $B$ decays~\protect\cite{Giorgi,Sakai} are 
in excellent agreement with the predictions of a global CKM 
fit~\protect\cite{Ligeti,CKMfitter}.}
\label{fig:CKM} 
\end{center}
\end{figure}

\begin{figure}[htb]
\begin{center}
\includegraphics[width=.45\textwidth]{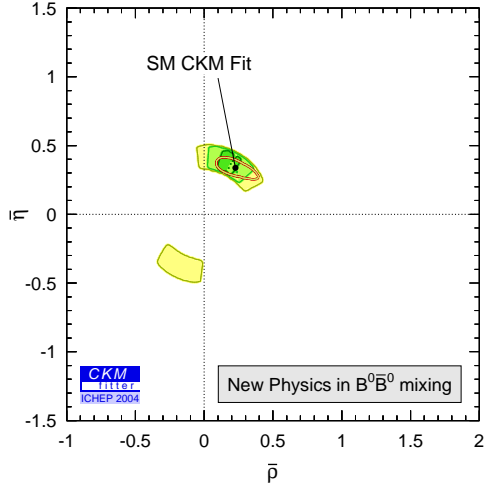}
\caption{
The agreement between CP violation 
measurements~\protect\cite{Giorgi,Sakai} and the 
global CKM
fit~\protect\cite{CKMfitter} leaves little room for new 
physics~\protect\cite{Ligeti}.}
\label{fig:NFP} 
\end{center}
\end{figure}

Putting together all the available information on quark flavour physics,
there is excellent consistency with the CKM model, as seen in
Fig.~\ref{fig:CKM}. Overall, there is little room left for new
physics~\cite{Ligeti}, as seen in Fig.~\ref{fig:NFP}.  Whatever there is,
it probably has a similar flavour structure similar to that of CKM, which
is a potential challenge for extensions of the Standard Model such as
supersymmetry~\cite{Masiero}.

However, there is still significant room for future progress, since the
experimental errors in many $B$ decay modes are still considerably larger
than the theoretical uncertainties~\cite{Ligeti}. Any larger discrepancy
would be evidence for new physics. Theoretical breakthroughs would be
needed to push the present theoretical errors much lower, but these decay
modes will not be theory-limited for a long time.

Before leaving quark flavour physics, it should be mentioned that
important consistency checks on the CKM model may be made in charm 
physics~\cite{Shipsey} and
using rare $K$ decays, particularly $K \to \pi \nu\bar{\nu}$ decays (in
both charged and neutral modes)~\cite{Patera}. Three candidate $K^+ \to
\pi^+ \nu\bar{\nu}$ decays have now been seen~\cite{E949}, and the rate
has been sinking towards the SM value, while progress is expected soon in
the search for $K_L \to \pi^0 \nu\bar{\nu}$ decay. The related $K_L \to
\pi \ell^+ \ell^-$ decays are also interesting for studies of CP
violation, and important groundwork for their interpretation is being
played by NA48 measurements of $K_S \to \pi^0 \ell^+ \ell^-$ and $\pi^0
\gamma \gamma$~\cite{NA48}. These constrain amplitudes that compete with
the direct CP-violating amplitude expected in the CKM model.

\section{Neutrino Masses and Oscillations}

Neutrinos provided the first confirmed physics beyond the Standard
Model~\cite{Barbieri,Langacker}, namely confirmed deficits in both the
atmospheric and solar neutrino fluxes. Another apparent deficit has been
observed by the LSND experiment at Los Alamos~\cite{LSND}, for which we
are awaiting confirmation from the MiniBooNE experiment at
Fermilab~\cite{MiniBooNE}.

\begin{figure}[htb]
\begin{center}
\includegraphics[width=.45\textwidth]{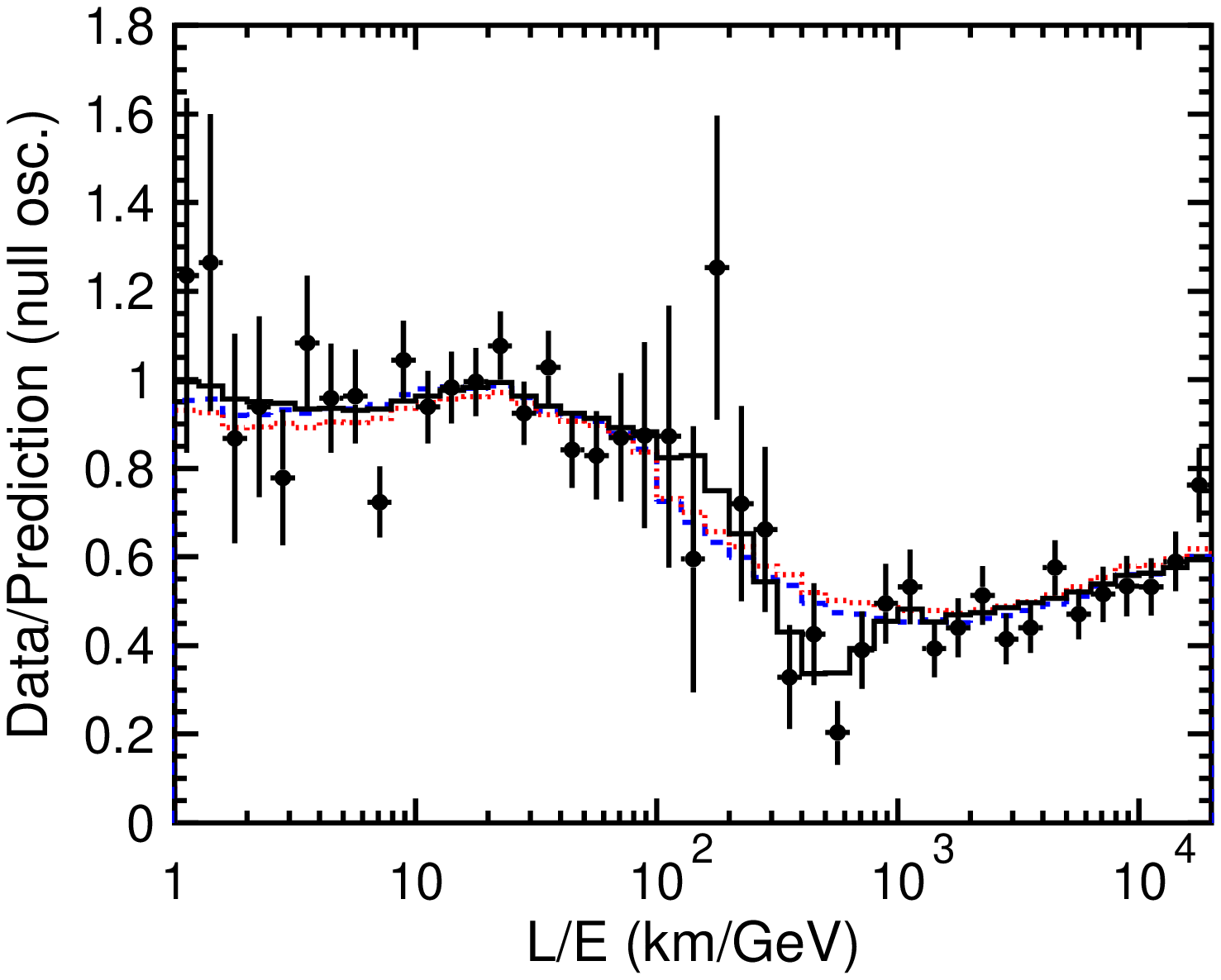}
\includegraphics[width=.32\textwidth,angle=-90]{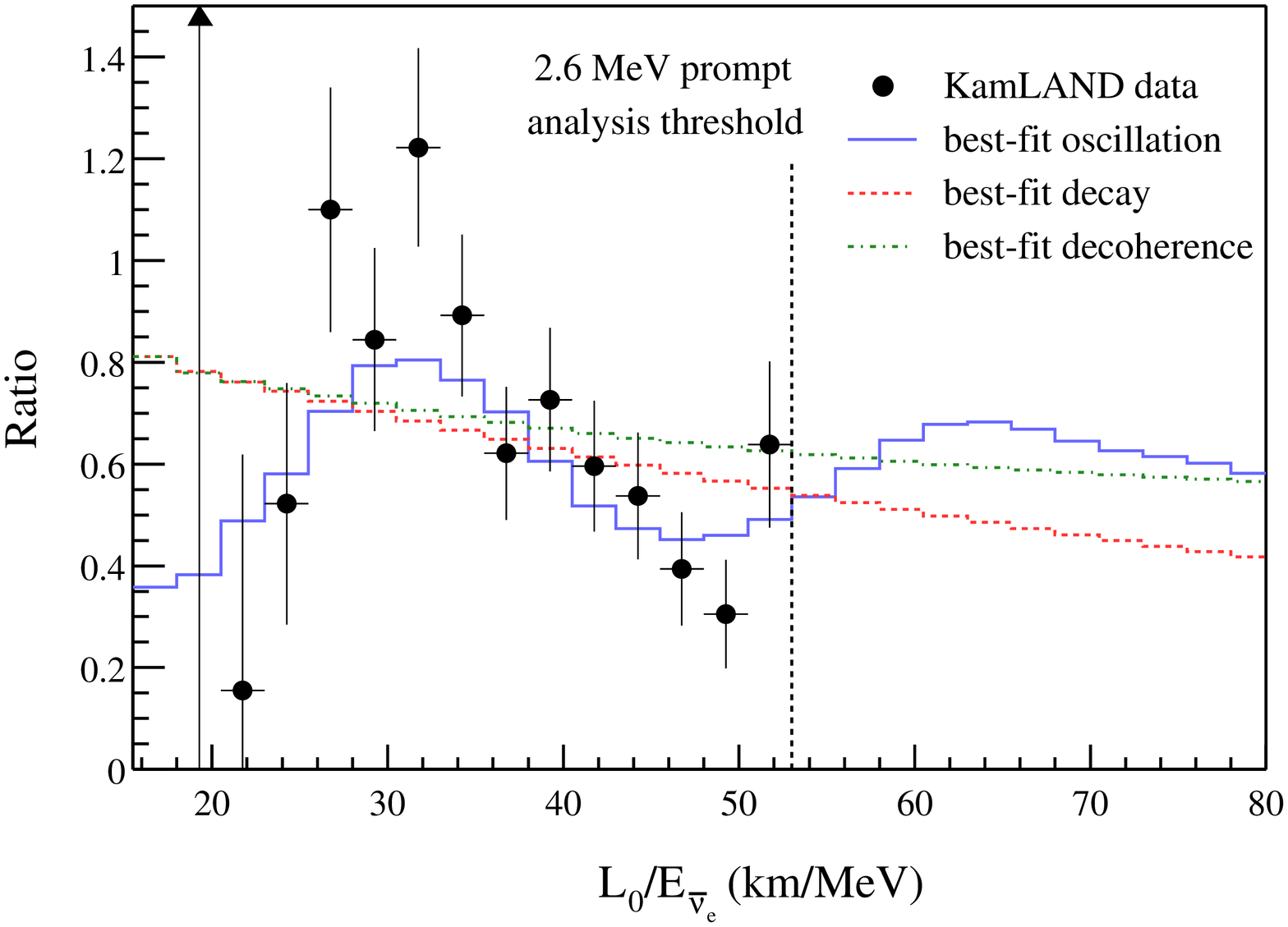}
\caption{
Chracteristic oscillatory patterns have been seen in atmospheric neutrino 
data (upper panel)~\protect\cite{SKosc} and in reactor neutrino 
data (lower panel)~\protect\cite{KLosc}.}
\label{fig:nuosc} 
\end{center}
\end{figure}

One of the most exciting developments in neutrino physics reported
here~\cite{Wang} has been the evidence for neutrino oscillation patterns
found by both the Super-Kamiokande~\cite{SKosc} and KamLAND~\cite{KLosc}
experiments, shown in Fig.~\ref{fig:nuosc}. Super-Kamiokande made a 
specific study of atmospheric
neutrino events with a good determination of $L/E$, and observed an
apparent oscillation dip. Rival hypotheses such as neutrino decay and
decoherence are disfavoured by $\chi^2$ (neutrino decay --- oscillation)
$= 11.4$ and $\chi^2$ (neutrino decoherence --- oscillation) $ =
14.6$~\cite{SKosc}. Evidence of similar strength has been reported by the
KamLAND experiment~\cite{KLosc} and there is also a hint of an oscillation
dip in K2K data~\cite{K2Knew}. Rivals to the oscillation hypothesis are
falling by the wayside.

Within the oscillation framework, the\\
new, higher-statistics KamLAND data are
strikingly~consistent~with~the 
Super- \\
Kamiokande and SNO data on solar
neutrinos~\cite{Wang}, with a similar mass-squared difference $\Delta m^2
\sim 8$ $\times 10^{-5}$ eV$^2$ and compatible mixing angles. Likewise,
accelerator neutrino data from K2K~\cite{McGrew} confirm Super-Kamiokande
on atmospheric neutrinos, with $\Delta m^2 \sim 2.5 \times 10^{-3}$
eV$^2$, and are also compatible with maximal mixing.

Interpreted in terms of neutrino oscillations, the atmospheric and solar
neutrino data indicate a (near) bimaximal mixing pattern completely
different from that found for quarks~\cite{Langacker}. Empirically, the
data are compatible with the relation $\Theta_{\odot} + \Theta_c = \pi/4$,
and a challenge for theorists is to figure out why this should be the
case~\cite{Rodejohann}. In addition to the two large mixing angles already
observed, there is a third mixing angle, $\Theta_{13}$, whose measurement
remains a challenge for future accelerator or reactor
experiments~\cite{McGrew}. The Holy Grail in neutrino oscillations is the
search for CP violation, but we now realize that this may have only an
indirect relation to cosmology via baryogenesis~\cite{notsame}.

Much still remains to do! Oscillations are made possible by differences in
neutrino masses, and we want to know whether they follow a normal or
inverted hierarchy. At a basic level, theorists speculate whether neutrino
masses are of Dirac or Majorana type. Many models are based on the
seesaw mechanism, but there is not a shred of evidence for it! Insight 
into these questions can be provided indirectly
by searches for neutrinoless double-$\beta$ decay and searches for the
violation of charged lepton numbers~\cite{Langacker}.

\section{Physics beyond the Standard Model}

The most pressing issue in physics beyond the Standard Model is breaking
electroweak symmetry (EWSB)~\cite{Barbieri}. This problem must be solved
below $ \sim 1$ TeV, and its solution would be a revolution in fundamental
physics that is a necessary basis for further theoretical speculations.
There are various hints of grand unification, such as the possible
unification of gauge couplings and also neutrino masses, but it is
difficult to test such unification ideas directly, and here I follow the
rapporteur~\cite{Barbieri} in focussing on EWSB.

If we adopt the calculability principle that the electroweak scale should
be calculable in terms of other physical mass scales, we are led to
consider models without quadratic divergences in the mass of the Higgs
boson. Then the two basic options are to invoke supersymmetry or to
interpret the Higgs boson as a pseudo-Goldstone boson~\cite{Barbieri}.
Supersymmetry has the supplementary advantages that it facilitates
unification of the gauge couplings~\cite{superGUT} and supplies a natural
candidate for astrophysical dark matter~\cite{EHNOS}. LEP data require
some fine-tuning of the supersymmetric model parameters, but (in contrast
to some others) I do not regard this as a severe problem~\cite{EOS}.

What are the alternatives to the default option of a light Higgs boson
accompanied by supersymmetry? One possibility considered has been to
re-examine the na\"{i}ve {\bf interpretation of the electroweak
data}~\cite{Chan}. Questions have been raised about the consistency of the
precision electroweak measurements, and it has been asked whether some
should be discarded, though, as already mentioned, the `discrepancy' 
between the direct and indirect limits on $m_H$ have now 
evaporated~\cite{Teubert,LEPEWWG}. As already remarked, heavy-flavour 
measurements 
alone tend to favour a heavier Higgs boson than is allowed by simple
supersymmetric models. Another possibility is that the global electroweak
fit should include contributions from {\bf higher-dimensional
operators}~\cite{BS} as well as the Higgs boson. Analyses of this type
find allowed corridors of parameter space extending to higher Higgs
masses. If the Higgs boson is indeed light, alternatives to supersymmetry
are provided by {\bf little Higgs models}~\cite{Schmaltz}, in which the
cancellation of one-loop quadratic divergences is ensured by an extra
`top-like' quark, more gauge bosons, and Higgs-like fields. At the other
extreme, theorists have been experimenting with {\bf Higgsless
models}~\cite{Rizzo}, in which WW scattering becomes strong at high
energies. Such models have problems with the precision electroweak data,
which may be alleviated in models formulated with extra dimensions, though
problems still remain.

\begin{figure}[htb]
\begin{center}
\includegraphics[width=.45\textwidth]{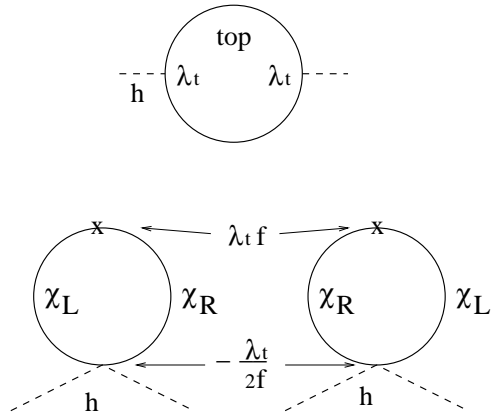}
\caption{
In `Little
Higgs' models, the quadratically-divergent top-quark loop contribution 
to $m_H^2$ is cancelled by
loops containing a heavier charge-2/3 quark~\protect\cite{Schmaltz}.}
\label{fig:LH}
\end{center}
\end{figure}

Among these various options, little Higgs models may be the most
interesting, so I discuss them in a bit more
detail~\cite{Schmaltz}. The Standard Model is embedded in a larger gauge 
group that is broken down to $SU(2) \times U(1)$, with the
light Higgs boson appearing as a pseudo-Goldstone boson. The quadratic
correction due to the top quark:
$$
\delta m^2_{H,top} (SM) \sim (115 \;{\rm GeV})^2
\left ( \frac{\Lambda}{400 \; {\rm GeV}} \right )^2
$$
is cancelled, as seen in Fig.~\ref{fig:LH}, by a new heavy $T$ quark with 
mass
$$
m_T > 2 \lambda_t f \sim 2 f
$$
where $f > 1$ TeV. As a result, the above quadratic divergence is softened
to a logarithmic one:
\begin{eqnarray}
\delta m^2_{H,top} (LH) & \sim &
\frac{6G_F m^2_t}{\sqrt{2} \pi^2} m^2_T \log \frac{\Lambda}{m_T} \nonumber \\
& \ge & 1.2 f^2 \nonumber
\end{eqnarray}
There is an upper bound on the mass of the new $T$ quark:
$$
M_T < 2 \; {\rm TeV} (m_H/200 \; {\rm GeV})^2
$$
There are also upper bounds on the expected new gauge bosons and Higgs bosons:
\begin{eqnarray}
M_{W'} & < & 6 \, {\rm TeV} (m_H/200 \, {\rm GeV})^2, \nonumber \\
M_{H^{++}} & < & 10 \, {\rm TeV}.\nonumber
\end{eqnarray}
More physics must appear above the $10$ TeV scale, so such a little Higgs
theory is not as complete as supersymmetry.

Is there any evidence from accelerator experiments for physics beyond the
Standard Model? The best claimant is the
measurement of $g_{\mu} - 2$ at BNL~\cite{BNL}, but its interpretation has 
been
clouded~\cite{Teubert} by uncertainties in the Standard Model
contribution calculated on the basis of the hadronic vacuum polarization
extracted from $e^+e^-$ data or $\tau$ decay
data~\cite{Davier}. The largest contributions to $g_{\mu} - 2$ and the 
most important
uncertainties come from low energies $< 1$ GeV.
In an important recent development, KLOE has released data~\cite{KLOE} 
that agree well
with those from CMD-2~\cite{CMD}, and the trend is
now to discard the $\tau$ data in the Standard Model evaluation of $g_{\mu}
- 2$~\cite{Teubert}, although this begs the question why
there is a $\sim 10\%$ discrepancy between the $\tau$ and $e^+e^-$  data
between the $\rho$ peak and $1$ GeV. However,
using the final BNL data and dropping the $\tau$ data, one now finds a
discrepancy with the Standard Model of~\cite{Teubert}
$$
a_{\mu}^{exp} - a_{\mu}^{SM} =
(25.2 \pm 9.2) \times 10^{-10},
$$
corresponding to a $2.7 \sigma $ effect. This is promising, but
insufficient to make a convincing claim for physics beyond the Standard 
Model. Nevertheless, it provides a significant constraint on many 
extensions of the Standard Model, such as supersymmetry.

Many direct searches for physics beyond the Standard Model were also
reported here, and I mention in particular the renewed search for squarks
and gluinos at the Tevatron collider using Run 2 data~\cite{Heinemann}.
General searches have now edged beyond the Run 1 limits, to 292 and 333
GeV for squarks and gluinos in one analysis, and a dedicated search for a
light sbottom squark excludes a significant new region of parameter
space~\cite{Heinemann}. Further progress from Run 2 is eagerly awaited.

We did not have a plenary talk about physics prospects for the LHC, but
some were presented in a parallel session~\cite{Barr}. Following discovery
(!?), several sparticle masses may be measured quite accurately at the LHC
in exclusive cascade decays~\cite{cascades}. Measurements of mass
differences will typically be limited by the detector performance to
errors of order $1\%$, whereas the error in the overall mass scale will be
dominated by that in the unknown missing energy to the order of
$10\%$~\cite{Barr}. New simulations indicate that it should be possible to
measure sparticle spins at the LHC, in some cases. Analayses have shown
how to measure the spin-1/2 nature of the $\chi_2$, and it has been shown
that one could also measure the scalar nature of the slepton at several
distinct points in parameter space~\cite{Barr}.

However, digging a supersymmetric signal out from the Standard Model
backgrounds will not be easy, and it is important to calculate the latter
as accurately as possible. There are plenty of necessary but unglamorous
NLO QCD calculations~\cite{Stirling} waiting to be done! Perhaps the new
string techniques will simplify them?

As we heard from the rapporteur~\cite{Barbieri}, the LHC has an important
location on the roadmap for exploring EWSB physics, and there are many
scenarios in which a 500-GeV LC could add significant physics value. If
naturalness is a good guide, the LHC will find signals of new physics. For
their full interpretation, a 500-GeV LC would often (though not always) be
very significant.

\section{A Few Remarks on String Theory}

String theory is not only very beautiful mathematics, but also provides
very powerful tools for field-theoretical calculations, e.g., for QCD as
discussed earlier~\cite{Stirling}. String theory has also aready solved
many major problems in Quantum Gravity, such as the extraction of sensible
results from perturbative calculations and the counting of black hole
states~\cite{Liu}.  String theory might explain the origin of the
Universe. It might replace our na\"{i}ve models of inflation based on an
elementary scalar field~\cite{EMNS}. However, our concern here is whether
it is relevant to particle physics. This is an open question, but there
are many exciting possibilities.The landscape of string vacua is very
rich~\cite{landscape}, and one or more of these states may be able to
describe all of particle physics. However, we are still lacking a
distinctive experimental signature. If they are ever discovered, extra
dimensions might provide such a `smoking gun', but this remains to be
seen.

\section{Particle Astrophysics and Cosmology}

Cosmology and high-energy astrophysics abound in important
problems~\cite{Binetruy} that only particle physics may be able to solve:
What is the Dark Matter? Is there Dark Energy? Are there ultra-high-energy
cosmic rays (UHECRs) beyond the GZK cutoff? Was there inflation? How did
the Universe begin?

According to the concordance model of cosmology, about $73\%$ of the
energy density of the Universe is unclumped, invisible Dark Energy, about
$23\%$ is clumped, invisible Dark Matter, and only about $4\%$ is visible
matter, with neutrinos an even smaller percentage~\cite{Binetruy}. This
concordance model has been supported by data on large-scale
structures~\cite{LSS}, high-redshift supernovae~\cite{hizSN} and
observations of the cosmic microwave background by WMAP et
al.~\cite{WMAP}. As we heard at this conference~\cite{Barbiellini}, 
gamma-ray
bursters are now emerging as another standard, normalizable candle capable
of measuring the geometry of the Universe~\cite{Ghirlanda}, and they tend
also to support this concordance model.

There has been continuing discussion here of particle candidates for the
cold Dark Matter~\cite{Binetruy}. These include the axion, WIMPs such as
the lightest supersymmetric particle (LSP) --- which might be either a
neutralino or a gravitino, the lightest Kaluza-Klein Particle (LKP) in
models with universal extra dimensions, and a superheavy (metastable)
`WIMPzilla' particle, which might have been produced at inflation, and
whose decays could be responsible for the UHECRs. Here I focus on the
neutralino WIMP possibility.

\begin{figure}[htb]
\begin{center}
\includegraphics[width=.45\textwidth]{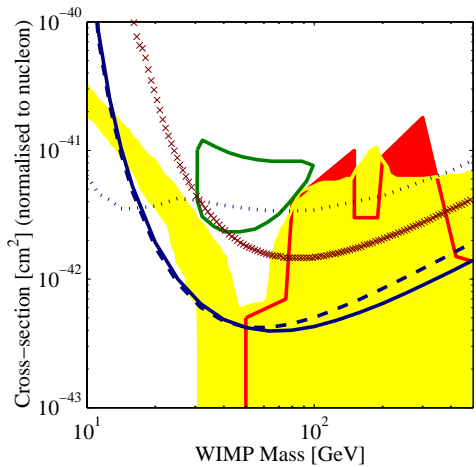}
\caption{
Upper limits on the cross section for spin-independent scattering between 
Dark Matter particles and nucleons compared with some theoretical 
models~\protect\cite{CDMS2}.}
\label{fig:CDMS}
\end{center}
\end{figure}

There has been significant recent \\
progress in the direct search for WIMP
Dark Matter, looking for elastic scattering on nuclei in low-background
experiments. The previous DAMA modulation signal is increasingly difficult
to reconcile with other experiments such as CDMS2~\cite{CDMS2}, which
currently has the best upper limit on Dark Matter scattering, as shown in 
Fig.~\ref{fig:CDMS}. This
experiment is already starting to reach into the range expected in
realistic models, and has good prospects for further improvement by a
factor $\sim 20$.

There have been discussions here of a possible signal of gamma rays coming
from neutralino annihilations near the galactic centre, since EGRET has
observed an apparent excess over standard cosmic-ray
calculations~\cite{Morselli,Sander}. However, there are large
uncertainties in the cosmic-ray background calculated with the standard
Galprop code, and there are also large unknowns in the prospective signal
normalization, so my advice is to `wait and see'.

A new upper limit was reported here by AMANDA on muons produced by
high-energy neutrinos emanating from neutralino annihilations inside the
Sun~\cite{AMANDA}, which is comparable with previous limits from BAKSAN,
MACRO and Super-Kamiokande. The IceCube experiment currently under
construction has good prospects to explore far into the range expected in
realistic models.

There are prospects that one of these Dark Matter experiments might be
able to rival the LHC in the search for supersymmetry~\cite{EFFMO}.

\section{Future Accelerators}

There are many complex issues related to future accelerators, many of
which are not technical. In some ways, the simplest question is what
accelerators we want to do the physics~\cite{Miller}. It seems that we
know enough to build interesting accelerators~\cite{Yokoya} and to do
experiments with the accelerators currently being proposed~\cite{Brau},
but it is less obvious that they will be able to answer all our
questions~\cite{Barbieri}. But we also have to figure out how we can
continue to involve a diversity of regions, and how we can ensure a
diversity of facilities. Certainly we shall need to work together to get
them approved, and also to build them~\cite{Dorfan}.  Finally, I should
like to add another question: How can we ensure access to new accelerators
from all qualified and interested physicists? We have recently had
instances of scientists from some countries being denied visas to attend a
major international conference, and IUPAP has announced that it will not
sponsor conferences where open attendance cannot be assured~\cite{Luth}.
What would be the attitude of IUPAP to the construction of a unique global
accelerator in a country that is not open to scientists from around the
world?

There are plenty of strong motivations for future colliders, since both
physics and cosmology lead us to expect strongly new physics at the scale
of 1 TeV~\cite{Miller}. As we have heard from the rapporteur here, the
physics case for the LHC has been made and accepted, and it will look into
the whole region where new physics can be expected. The physics case for a
TeV International Linear Collider (ILC) has also been made, and the
physics cases for CLIC (and perhaps a larger hadron collider) will be
understood better following the results from the LHC (and perhaps a TeV
ILC)~\cite{Miller}.

\begin{figure}[htb]
\begin{center}
\includegraphics[width=.45\textwidth]{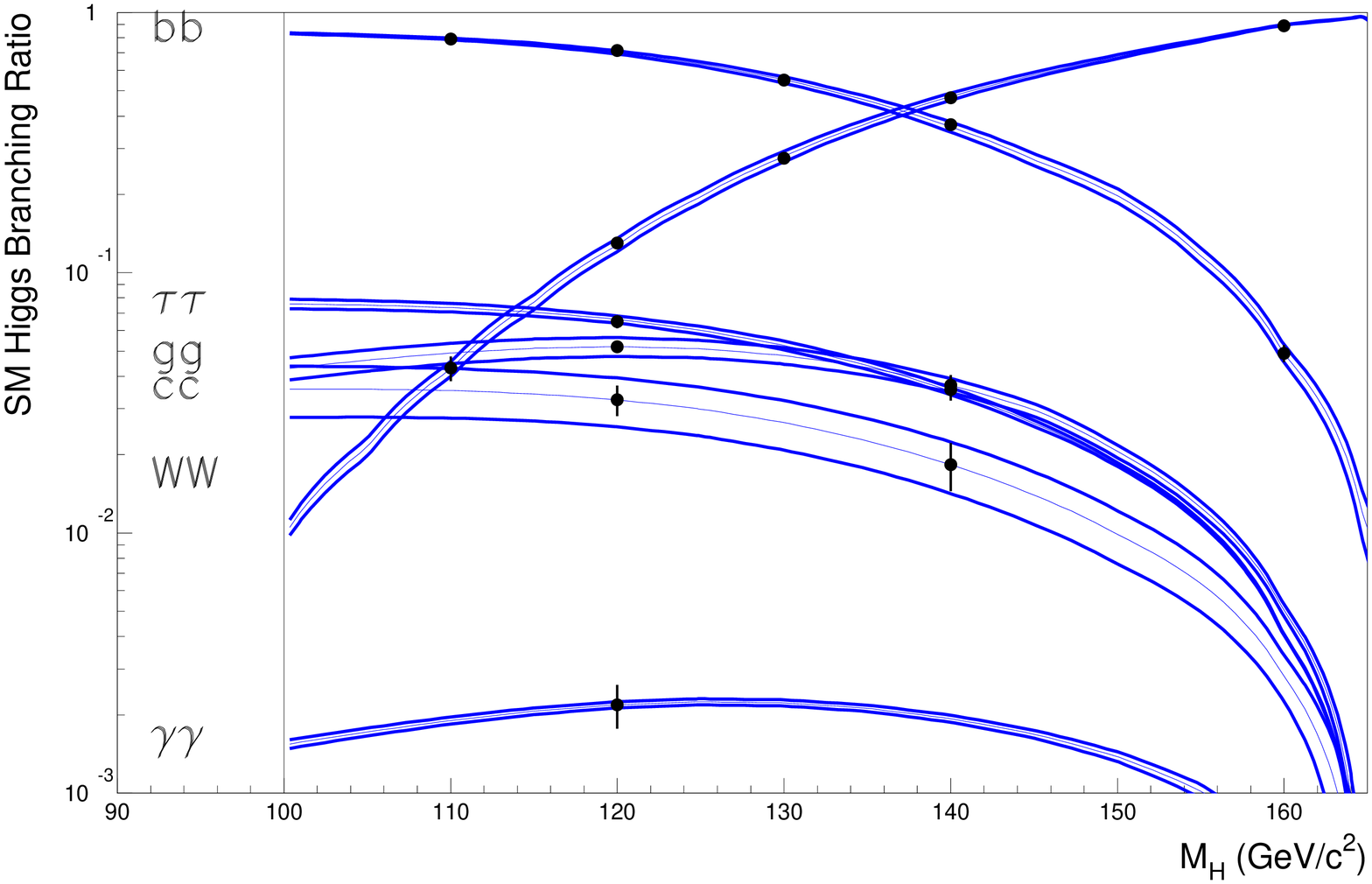}
\caption{
The decay branching ratios of a light Higgs boson could be measured very 
accurately with the ILC~\protect\cite{Miller}.}
\label{fig:ILCHiggs} 
\end{center}
\end{figure}

The principal tasks for a TeV ILC are well understood~\cite{Miller}. It
should measure $m_t$ with an accuracy $< 100$ MeV. If there is a light
Higgs boson of any kind, the ILC should pin it down: What is its precise
mass and does it have Standard Model couplings? As seen in
Fig.~\ref{fig:ILCHiggs}, the decays of a light Higgs - if it exists -
could be measured very accurately at the ILC. Moreover, if there are extra
light particles, such as those predicted by supersymmetry, the ILC would
also measure their masses and properties very well. On the other hand, if
the LHC sees nothing new below $\sim 500$ GeV, the ILC should look for
indirect signatures, such as those of a new $Z'$ or a high-mass $W_L W_L$
resonance.

One of the most dramatic developments here has been the
announcement~\cite{Dorfan} of the ITRP recommendation that the ILC be
based on superconducting $rf$ technology~\cite{ITRP}, and its endorsement
by ICFA. We have been told that the features of the superconducting
technology that tipped the balance in its favour follow in part from the
low $rf$ frequency. The hope is now that the final design of the ILC will
be developed by a team drawn from the combined warm and cold linear
collider communities, taking full advantage of the experience and
expertise of both~\cite{Yokoya}.

Beyond the sub-TeV ILC, we believe we shall need a multi-TeV linear
collider~\cite{Paris}, for which the only current contender is CLIC with a 
nominal
design energy of 3 TeV~\cite{Miller}. CLIC is based on a novel two-beam
accelerator concept, rather than conventional klystrons. Two CLIC test
facilities have already operated successfully, and third CLIC test
facility has now entered operation. It is planned to pass all the
showstopping R1 and R2 R\& D milestones, by 2009 if sufficient resources
can be found, about five years after the ILC.

\begin{figure}[htb]
\begin{center}
\includegraphics[width=.45\textwidth]{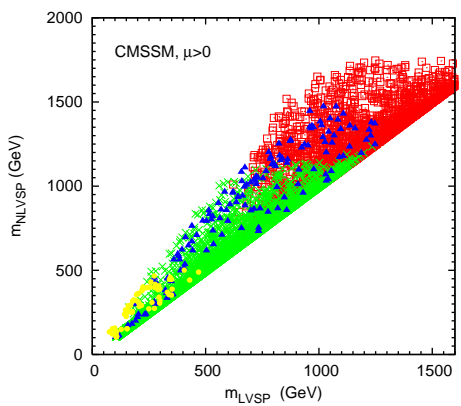}
\end{center}
\caption{
Scatter plot of the masses of the lightest visible
supersymmetric particle (LVSP) and next-to-lightest visible
supersymmetric particle (NVSP) in the constrained MSSM with universal 
input scalar masses~\protect\cite{EOSS}: similar results were found in 
other supersymmetric scenarios. The blue points have an appropriate 
Dark Matter density, the green points should be accessible to 
the LHC, and the yellow points might be detectable directly in Dark 
Matter scattering experiments.} 
\label{fig:scatter}
\end{figure}

Although the physics agenda for CLIC cannot yet be completely defined, I
am convinced that it will be very rich. Consider just the example of
supersymmetry. The threshold for supersymmetry may be within the reach of
the sub-TeV ILC, but, as seen in Fig.~\ref{fig:scatter}, this cannot be 
guaranteed~\cite{EOSS}. Even if some
supersymmetric particles do have masses low enough to be observed at the
ILC, most of the spectrum will lie beyond its reach, and CLIC would be the
natural machine to study it~\cite{Bench,CLICphys}.

\section{Summary of the Summary}

As we have heard at this conference, QCD is becoming ever more
quantitative, but there remain qualitative puzzles, particularly in
spectroscopy. Electroweak theory and experiment suggest new physics at the
TeV scale, very likely a Higgs boson and perhaps many other particles in
addition. Flavour physics is also becoming more quantitative, and the
highlight of this conference has been that the CKM is looking better and
better all the time. We have also seen here that neutrinos really do
oscillate. The symbiosis with cosmology is growing, and the LHC is on its
way. There are many good ideas for future accelerators, and the ITRP has
done its work in choosing a technology for the ILC. Now is the time for us
to get back to our own work!


\begin{thebibliography}{99}
\bibitem{Danilov}
However, M.~Danilov, talk in the `Hadron Spectroscopy' parallel session 
at this conference, reported negative results from a low-energy search in 
secondary interactions produced inside the Belle detector.
\end{thebibliography}
\end{document}